\documentstyle[12pt]{article}

\def\reset{\setcounter{equation}{0}}

\def\GeV{\,{\rm GeV}}

\def\MeV{\,{\rm MeV}}
\def\sec{\,{\rm sec}}

\def\yr{\,{\rm yr}}
\def\rcm{\,{\rm cm}}

\def\eV{{\,\rm eV}}

\def\cmm2{{\,\rm cm^{-2}}}
\def\cm2{{\,{\rm cm}^2}}
\def\cmm3{{\,{\rm cm}^{-3}}}
\def\gcmm3{{\,{\rm g\,cm^{-3}}}}

\def\mpl{{m_{\rm Pl}}}
\def\mpp{{m_{\rm Pl,0}}}

\def\g{\tilde g}
\def\R{{\cal R}}

\def\la{\mathrel{\mathpalette\fun <}}
\def\ga{\mathrel{\mathpalette\fun >}}
\def\fun#1#2{\lower3.6pt\vbox{\baselineskip0pt\lineskip.9pt
  \ialign{$\mathsurround=0pt#1\hfil##\hfil$\crcr#2\crcr\sim\crcr}}}

\begin{document}
\pagestyle{empty}
\begin{center}
{\Large \bf DYNAMICAL SOLUTIONS TO THE}\\
\medskip
{\Large \bf HORIZON AND FLATNESS PROBLEMS} \\

\vspace{.2in}
Yue Hu,$^1$ Michael S. Turner,$^{2,3}$ and Erick J.~Weinberg$^1$\\

\vspace{.2in}
$^1${\it Department of Physics, Columbia University,
New York, NY  10027}\\

\vspace{0.1in}
$^2${\it Departments of Physics and of Astronomy \& Astrophysics\\
Enrico Fermi Institute, The University of Chicago, Chicago, IL  60637-1433}\\

\vspace{0.1in}
$^3${\it NASA/Fermilab Astrophysics Center,
Fermi National Accelerator Laboratory, Batavia, IL  60510-0500}\\

\end{center}

\vspace{.3in}

\centerline{\bf ABSTRACT}
\vspace{0.15in}

We discuss in some detail the requirements on an early-Universe model that
solves the horizon and flatness problems during the epoch of
classical cosmology ($t\ge t_i\gg 10^{-43}\sec$).
We show that a dynamical resolution of
the horizon problem requires superluminal expansion (or very close
to it) and that a truly satisfactory resolution of
the flatness problem requires entropy production.
This implies that a proposed class of adiabatic
models in which the Planck mass varies by
many orders of magnitude cannot fully resolve the flatness problem.
Furthermore, we show that, subject to minimal assumptions, such
models cannot solve the horizon problem either.  Because superluminal
expansion and entropy production are the two generic features
of inflationary models, our results suggest that
inflation, or something very similar, may be the only dynamical
solution to the horizon and flatness problems.

\newpage
\pagestyle{plain}
\setcounter{page}{1}
\newpage

\section{Introduction}
\reset

As successful as the standard cosmology is, it has
two well known shortcomings:  the horizon
and flatness problems \cite{peebles,guth}.
These shortcomings do not indicate any logical inconsistency,
but rather that in the standard cosmology
the present state of the Universe depends strongly upon
the initial state---a feature that many consider undesirable.
It is possible that these shortcomings do
not require an explanation.  Penrose has suggested that
there may be a law of physics
that governs the initial state of the Universe \cite{penrose}.
Or perhaps the required initial state for the classical epoch of
cosmology is dictated by the outcome of the quantum-gravity
era.  Here we focus on dynamical solutions to these
problems during the era when gravity can be treated
classically ($t\ge t_i \gg 10^{-43}\sec$).

Guth's inflationary Universe paradigm provides an elegant solution
involving the microphysics of the very early Universe
($t\sim 10^{-34}\sec$) \cite{guth}.  While Guth's original model
based upon a first-order symmetry-breaking phase transition
did not work, many viable implementations of inflation now exist
\cite{inflation}.  All involve two key
elements: a period of superluminal expansion (driven by scalar-field
potential energy) and massive entropy production (conversion of that
energy to radiation).  We show that both
features are essential for a true resolution of these problems:
Superluminal expansion (or very close to it)
is required to solve the horizon problem,
while one cannot account for the flatness of
the present Universe for arbitrary initial curvature without entropy
production.

Of course, one might be less ambitious and try to solve only the
horizon problem.  Indeed, it has been proposed that an adiabatic
scenario based on a time-varying Planck mass
might provide a solution to the horizon
problem \cite{zee,soft,janna}.  We examine this
proposal and show, with minimal assumptions (essentially
positivity of energies), that in the context of scalar-tensor
theories a varying Planck mass cannot lead
to a solution of the horizon problem either.

In Sec.~2 we discuss the horizon and flatness problems.  We show why
superluminal expansion (which we define more precisely) and entropy
production are essential ingredients of any solution to these
problems.  We describe how the horizon and flatness
problems are solved in inflationary
scenarios in Sec.~3.  Scalar-tensor theories with a time-varying
Planck mass are discussed in Sec.~4, and we show that they cannot
solve the horizon problem without entropy production \cite{extended}.
Section~5 contains our concluding remarks.

\section{The Cosmological Problems}
\reset
\subsection{The horizon problem}

The horizon problem involves the smoothness---that is, homogeneity
and isotropy---of the presently
observed Universe (size $\sim H_0^{-1}\sim 10^{28}\rcm$).
The uniformity of the cosmic background
radiation indicates that this volume was
smooth at $t \sim 300,000\yr$ ($T\sim 0.3\eV$),
when matter and radiation decoupled; further,
the successful predictions of primordial nucleosynthesis
are evidence that it was smooth at least as early as $t \sim
1\sec$ ($T\sim 1\MeV$).  However, in the standard cosmology the
distance that a light signal can travel by such early times, $d_{\rm HOR}(t) =
R(t)\int_{t_i}^t du/R(u)$, is much smaller than the
size of the presently observed Universe at that time,
$d_U(t) \sim R(t) H_0^{-1}/R_0$; this precludes causal physics
operating at early times from accounting for the smoothness.
(Here $R(t)$ is the cosmic-scale factor, $H\equiv \dot R/R$
is the expansion rate, and a subscript 0 denotes the present epoch.
Throughout, we assume that recombination and last-scattering took place at
a temperature $T\sim 0.3\eV$, as expected.  None
of our conclusions change if the Universe remained ionized
so that these events occurred later.)

A necessary (though not sufficient) condition to solve the horizon
problem is
\begin{equation} \label{eq:horcondx}
d_{\rm HOR}(t_S) =
R(t_S)\int_{t_i}^{t_S} dt^\prime /R(t^\prime ) > R(t_S)H_0^{-1}/R_0;
\end{equation}
where $t_S$, the time by which the horizon problem is solved,
must be less than $1\sec$, since
the presently observed Universe was already smooth by then.
Now note that if  $R(t)\propto t^n$ with $n<1$, then
$d_{\rm HOR}(t_S) \approx t_S/(1-n) = [n/(1-n)]H(t_S)^{-1}
\approx H(t_S)^{-1}$.  Thus, in the standard cosmology, where
$n=1/2$ in the radiation-dominated era ($t\la
10^{11}\sec$) and $n=2/3$ in the matter-dominated era
($t\ga 10^{11}\sec$), the right-hand side of condition
(\ref{eq:horcondx}) is {\it greater} than the left by a factor of
$(10^{15}\sec/ t_S)^{1/2}$ for any $t_S$ in the radiation-dominated
era and by a factor of $(3\times 10^{17}\sec /t_S)^{1/3}$
in the matter-dominated era.  Hence, condition (\ref{eq:horcondx})
cannot be satisfied.

It is instructive to recast the horizon-problem-solving condition.
Suppose first that at early times there was
``standard'' evolution (i.e., $R$ growing no faster than $t^n$
with $n$ less than, and not too close
to, unity---say $n < .99$) up to some time $t_1$, and that the
evolution was then nonstandard throughout
the interval $t_1 < t < t_S$.  Now write the integral in
Eq.~(\ref{eq:horcondx}) as the sum of the contribution
from $t_i$ to $t_1$ and that from $t_1$ to $t_S$.  The
first integral is dominated by the upper end of
the integration range and is equal,
up to factors of order unity, to $(R_1 H_1)^{-1}$, where subscript 1
refers to time $t_1$.  If $R$ grows faster than $t^n$, with $n$
greater than, and again not too close to, unity for $t_1<t < t_S$,
then the second integral is dominated by the lower end of its
integration range and is approximately equal to the first
integral.  Then, apart from factors of order unity,
Eq.~(\ref{eq:horcondx}) becomes
\begin{equation}\label{eq:rhcondition}
(R_1H_1)^{-1} > (R_0H_0)^{-1}.
\end{equation}
[If there were several alternating periods of standard and nonstandard
evolution of the scale factor, we divide the integration
range correspondingly, and
find that the integral in Eq.~(\ref{eq:horcondx}) is given by a sum of
terms of the form $[R(t_a)H(t_a)]^{-1}$, where the $t_a$ are the times
at which the periods of standard evolution terminate.  In general, one
of these terms will dominate; denoting as $t_1$ this time $t_a$,
we again obtain Eq.~(\ref{eq:rhcondition}).]

Condition (\ref{eq:rhcondition}) says that the comoving Hubble volume
at some very early time must grow large enough to
contain the present Hubble volume.  Since $RH=\dot R$, it further
implies that ${\dot R}_0 > {\dot R}_1$, which requires
that $\ddot R >0$ sometime between $t_1$ and $t_S\la 1\sec$.
Such expansion is often referred to as ``superluminal.''

Let us clean up a several minor points before going on.
First, suppose that there was no early period of ``standard evolution''
following the quantum-gravity era.  If the horizon distance at the end
of the quantum-gravity era, $d_{\rm HOR}(t_i)$, was of the order of
$H_i^{-1}$, we simply set $t_1 =t_i$ in Eq.~(\ref{eq:rhcondition}).
If instead the size of causally connected regions at
time $t_i$ was actually significantly greater than $H_i^{-1}$
and large enough to solve the horizon problem, then
from our perspective the horizon problem was solved
during the quantum-gravity era.

The second point involves
the kinematical motivation for the term superluminal.
Superluminal expansion might be most naturally
defined as that where {\it any} two comoving points eventually
lose causal contact; i.e., expansion so rapid that
$\int_t^\infty du/R(u)$ converges.  For $R\propto t^n$, this
corresponds to $n>1$, and hence $\ddot R >0$, the
definition adopted here.

Next, we note that there is a small loophole in the  arguments going
from Eq.~(\ref{eq:horcondx}) to
Eq.~(\ref{eq:rhcondition}).  If $R$ is very close to linear in time,
the estimates we made for the integral in Eq.~(\ref{eq:horcondx}) are
incorrect, and, {\it in principle,} the horizon problem can be solved
without superluminal expansion.  However, we stress that achieving
this requires a rather strictly constrained behavior for a long time.
For example, condition (\ref{eq:horcondx}) can be satisfied with $R
\propto t^{1-\epsilon}$ (and hence $\ddot R < 0$)
and $t_S \la 1\sec$ provided $\epsilon \la
10^{-7}$. While not superluminal, such behavior might well be termed
``almost superluminal''---or at least, nonstandard.

Finally, there is an apparent paradox in discussing
the horizon problem in the context of a Friedmann-Robertson-Walker
(FRW) model, which, by assumption, is isotropic and homogeneous.
Consideration of the most general cosmological solutions,
which are not homogeneous nor even fully
classified, is very difficult
though some attempts have been made \cite{belinsky}.  A more modest
approach to addressing this paradox is to consider a perturbed
FRW model that is {\it more} inhomogeneous than ours (which
is very easy to do since our Universe is so very smooth).
In this case, the FRW equations apply to lowest
order,  and it makes sense to discuss whether the level
of inhomogeneity can be reduced to a level consistent
with that observed.  Within this perturbative framework
the horizon problem
as usually discussed is correctly stated.  One can also consider
the isotropy issue alone by appealing to the homogeneous,
but nonisotropic, Bianchi models which have tractable field equations
\cite{aniso}.

\subsection{The flatness problem}

The flatness problem involves the observation that $\Omega$, the ratio
of the energy density of the Universe to the
critical density, is close to unity today in spite of the fact
that $|\Omega -1|$ grows as a power
of the scale factor.  In the FRW cosmology the expansion rate is governed by
the Friedmann equation,
\begin{equation}\label{eq:frweq}
H^2 = {8\pi G\rho \over 3} - {k\over R^2},
\end{equation}
where $\rho$ is the total energy density,
$\rho_{\rm crit} = 3H^2/8\pi G$ is the critical density,
and $R_{\rm curv} = R(t)|k|^{-1/2}$ is the curvature radius.
Eq.~(\ref{eq:frweq}) can be rewritten as
\begin{equation}
\Omega -1 = \left[\left(8\pi G\rho \over 3\right)
   \left(R^2\over k\right) -1 \right]^{-1}.
\end{equation}
For $\Omega$ close to unity,  $|\Omega -1|$
grows as $R^m$ with $m=2$ during the radiation-dominated
era and $m=1$ during the matter-dominated era.
Since the curvature radius is proportional to $|\Omega -1|^{-1/2}$,
\begin{equation} \label{eq:omega}
R_{\rm curv} = R(t)|k|^{-1/2} = {H^{-1}\over |\Omega -1|^{1/2}},
\end{equation}
it decreases relative to the Hubble radius $H^{-1}$ and the Universe
becomes ``less flat'' with time.  This means that the Universe
must have been very flat at the initial epoch
in order that $R_{\rm curv} \ga H_0^{-1}$ today:
$|\Omega_i -1| =(R_{\rm curv}H)^{-2} \ll 1$.
(In alternative theories of
gravity, we take the 3-curvature to be
$6k/R(t)^2$ and use Eq.~(\ref{eq:omega})
as the definition of $\Omega$.  This definition can be
used for any metric theory of gravity.)

The resolution of the flatness problem is
closely related to that of the horizon problem.
{\it Suppose} that the horizon problem is solved in the post
quantum-gravity era; i.e., that Eq.~(\ref{eq:rhcondition}) is satisfied with
$$(R_1H_1)^{-1} = \beta (R_0H_0)^{-1},$$
and $\beta\ge 1$.  It then follows by simple algebra that
\begin{equation} \label{eq:proof}
|\Omega_0-1| = |\Omega_1-1|/\beta^2.
\end{equation}
The deviation of $\Omega$ from unity is reduced
by the square of the factor by which the horizon problem is solved,
and thus $|\Omega -1|$ need not be initially set to a very small
value to insure that $|\Omega_0-1|$ is small today.

As we shall discuss later, it is possible
that the sequence of events that solves the horizon
problem (e.g., inflation) is prevented
from occurring by large curvature (e.g., the Universe recollapses
before it can inflate).  The point here is simply that, {\it if}
the horizon problem as formulated above is solved
after the emergence from the quantum-gravity era, {\it then}
one need not invoke an unnaturally flat initial state to
account for the flatness of the present Universe.

\subsection{Entropy considerations}

It is instructive to reformulate the flatness problem in terms of
entropy.  By a very wide margin most of the entropy in the Universe
exists in the form of radiation, today in the cosmic backgrounds of
$3\,$K photons and $2\,$K neutrinos, and at very early times in a
thermal bath of all particle species much less massive than the
temperature.  If the Universe remains close to thermal equilibrium,
the radiative entropy per comoving volume $S \propto g_*(RT)^3$ is
constant and the expansion is said to be adiabatic.  Here $g_*$ counts
the total number of effectively massless (mass $m\ll T$) degrees of
freedom; although $g_*$ varies during the evolution of the Universe we
can neglect this variation in our considerations\rlap.\footnote{A
closer analysis shows that the variation in $g_*$ does not affect any
of our conclusions unless it changes by a factor of
greater than $10^{180}$.}

Our present Hubble volume contains an entropy $S_0 \sim H_0^{-3}T_0^3
\sim 10^{88}$; for adiabatic expansion this is always the entropy
within the comoving volume corresponding to the present Hubble volume
since $RT=\,$const by adiabaticity.  In the standard cosmology,
the entropy within a horizon volume during the radiation-dominated
era, $S_{\rm HOR}(t) \sim H^{-3}T^3 \sim (\mpl /T)^3$, was
much smaller than this at early times, which is another way of stating
the horizon problem.

The curvature of the
Universe can also be characterized by the entropy contained within a
curvature volume: $S_{\rm curv} \sim R_{\rm curv}^3T^3=
|k|^{-3/2}(RT)^3$.  Since the curvature radius at present is
comparable to, or greater than, the Hubble radius, $S_{\rm curv}$ must
be at least $10^{88}$.
For adiabatic expansion, this is always the entropy within a curvature volume.
This allows us to express the size of the
curvature radius in terms of the temperature
at any epoch:
\begin{equation} \label{eq:obstacle}
R_{\rm curv}(t)  = S_{\rm curv}^{1/3} T^{-1}
\ga 10^{29}T^{-1}
\end{equation}
If furthermore $ H \ga T^2/\mpl$, as we usually expect, then
\begin{equation} \label{eq:ihatelatex}
R_{\rm curv}(t) H(t) \ga  10^{29} \left({T\over \mpl}\right).
\end{equation}
The upshot of Eqs.~({\ref{eq:obstacle},\ref{eq:ihatelatex})
is that in an adiabatic scenario it is not possible to
evolve to a universe as flat as ours if the curvature radius
at the initial epoch was smaller than $10^{29}T_i^{-1}$.

In the previous subsection, we described the flatness of the
present-day Universe in terms of the small size of $|\Omega -1| =
(R_{\rm curv}H)^{-2}$.  From that perspective, the
flatness problem is that, in the
standard cosmology, $R_{\rm curv}$ must have been
very much larger than the Hubble radius at early times.  From
the entropy
perspective, the puzzle is why the entropy within a curvature volume is so
extraordinarily large.  While one might envision an adiabatic scenario that
resolved the first formulation of the flatness problem, such a scenario
cannot change the entropy within a curvature volume and hence cannot
explain the enormous size of $S_{\rm curv}$.  Since the
entropy within a curvature volume is just another way of
specifying the initial curvature of an adiabatic universe, in an adiabatic
model the present-state of the Universe will always be
sensitive to the initial curvature:  curvature corresponding
to $S_{\rm curv}\la 10^{88}$ can never lead to a universe
that today is as flat as ours.  It is for this reason that we argue
that a truly satisfactory solution of the flatness problem cannot be
obtained within an adiabatic scenario.

We showed previously that the horizon and flatness problems were
linked:  Any scenario that solves the horizon problem
automatically accounts for the smallness of $|\Omega -1|$
today.  This of course also applies to an adiabatic scenario, and
reveals a further problem.  According to Eq.~(\ref{eq:ihatelatex})
the smallness of $|\Omega_0 -1|$ requires either,
(i) $ R(t_1) H(t_1) $ was very large, or
(ii) $T_1/\mpl$ was very small.  The horizon-problem-solving condition,
 Eq.~(\ref{eq:rhcondition}), does not allow the former.  Hence, {\it any}
adiabatic solution to the horizon problem has the
additional drawback that it
must take place at an unnaturally small temperature, many orders of
magnitude below the Planck mass.  (In Section 4, we show that
in scalar-tensor theories the horizon problem cannot be solved adiabatically
at all.)

One final point; we have used adiabaticity to mean constant
entropy per comoving three volume.  There have been attempts
in higher-dimensional theories of gravity to increase the
entropy per comoving three volume while maintaining adiabaticity
in the higher dimensional theory.  In such theories adiabaticity
refers to constant entropy per total comoving volume; the shrinking
of the volume of the extra dimensions causes the entropy
per comoving three volume
to increase.  Thus far this approach has not proven successful
\cite{extra}; further, in our terminology these scenarios
are not adiabatic (though this is largely a matter of semantics).

\section{How Inflation Resolves the Problems}
\reset

In this Section we illustrate how superluminal expansion
and entropy production solve the horizon and flatness
problems in the context of inflationary models.
The Universe today is matter
dominated with $\rho_{\rm matter} \sim 10^4\rho_{\rm rad}\sim
10^4 T_0^4$, so that $H_0 \sim
10^2T_0^2/\mpl$.  If at time $t_1$ the Universe was
radiation dominated, then $H_1 \sim T_1^2/\mpl$;
the horizon/flatness-problem-solving condition,
Eq.~(\ref{eq:rhcondition}), becomes
\begin{equation}  \label{eq:hsc}
{\mpl \over T_1S_1^{1/3}} \ga 10^{-2} {\mpl \over T_0S_0^{1/3}};
\end{equation}
where $S\propto (RT)^3$ is the entropy per comoving
volume.\footnote{In Eq.~(\ref{eq:hsc}) we need a
lower bound to $H_1$; the presence of additional forms
of energy density (which we assume to be positive)
or of negative curvature would only serve
to increase $H_1$, thus giving an even stronger condition.  Positive
curvature could decrease $H_1$; however, a closer analysis
shows that this decrease cannot be significant.  In alternative
theories of gravity there may be additional terms that might appear
to be able to reduce $H_1$; we will address this point, in the context
of a theory with a variable Planck mass, below.}
Equation~(\ref{eq:hsc}) can be satisfied if there is large-scale
entropy production:  $S_0/S_1 \ga (10^{-2}T_1/T_0)^3$.
This is the strategy underlying the inflationary solution:
A period of superluminal expansion (nearly exponential)
is followed by a reheating event which increases the
entropy by a large factor.

It is quite natural that superluminal
expansion should be followed by entropy production.
In the FRW cosmology the energy
density of a fluid with equation of state $p=\gamma \rho$ evolves as
$\rho \propto R^{-3(1+\gamma )}$, and if it dominates the energy
density the scale factor evolves as $t^{2/3(1+\gamma
)}$.  Superluminal expansion requires that
$\gamma < -1/3$, which implies that $\rho$ decreases more slowly
than $R^{-2}$.  During the superluminal phase, the energy density of
the ``fluid'' that drives inflation increases as $R^2$ (or faster)
relative to the radiation energy density, and the Universe
supercools to a very low temperature.  In order that the
Universe become radiation dominated once again the fluid
driving inflation must ``decay'' into radiation, and the
decay of this fluid into radiation
``reheats'' the Universe, thereby increasing the entropy by a large amount.

Let us now be more specific.  Suppose that the vacuum
energy that drives inflation is $V_0 \equiv {\cal M}^4$ and
that the dynamics of the inflation are such that it takes
a time $NH_I^{-1}$ for the vacuum energy to decay, after which
the Universe reheats to a temperature $T\sim {\cal M}$
(perfect conversion of vacuum energy to radiation).
Here $\cal M$ is the energy scale of inflation and $H_I\sim
{\cal M}^2/\mpl$ is the Hubble constant during inflation.  It
is convenient to write the Friedmann equation in a
slightly different form:
\begin{equation} \label{eq:frwentropy}
H^2 = {8\pi G \over 3}\left( {g_*\pi^2 \over 30}T^4 +V_0\right)
- {k\over R^2} \sim {T^4\over \mpl^2} - {{\rm sign}(k)T^2\over
S_{\rm curv}^{2/3}} + {{\cal M}^4 \over \mpl^2},
\end{equation}
where for simplicity we ignore the energy density in
nonrelativistic matter (which is negligibly small at early times) and
all the purely numerical factors.
As before, $S_{\rm curv}$ is the entropy within a curvature
radius before inflation, which increases greatly after inflation
due to the entropy production associated with reheating.

First, suppose that the curvature of the Universe
is not ``large;'' specifically that, $S_{\rm curv}\ga
\mpl^3 /{\cal M}^3$.  In this case, when the vacuum-energy
density starts to exceed the radiation-energy density,
at a temperature $T_1\sim {\cal M}$, and the Universe
begins to inflate, the
curvature term is still small compared to these terms
(at the beginning of inflation $\Omega$ is still
close to unity).  The Universe then grows in size by a
factor of $e^{N}$; since the temperature after inflation
is about the same as it was before inflation the entropy
per comoving volume increases by a factor of $e^{3N}$.

How large must $N$ be to solve the horizon and flatness
problems?  A Hubble-radius-sized patch at the beginning
of inflation, $H_I^{-1}\sim \mpl /{\cal M}^2$, grows by
a factor of $e^N$ by the end of inflation.  By today it
has grown by another factor of ${\cal M}/T_0$, since we assume
that the expansion is adiabatic after inflation.  This patch will
be larger than the present Hubble radius if
\begin{equation}\label{eq:nn}
N\ga 68 + \ln ({\cal M}/\mpl ).
\end{equation}

Next, the flatness problem; the entropy within a curvature
radius after inflation increases by a factor of $e^{3N}$
and is thus greater than $e^{3N} (\mpl /{\cal M})^3$
(since by assumption $S_{\rm curv} \ga \mpl^3/{\cal M}^3$).
Solving the flatness problem requires that this number
be greater than about $10^{88}$, or
\begin{equation}
N \ga  68 + \ln ({\cal M}/\mpl ) ,
\end{equation}
which is precisely the  condition for solving the horizon
problem.  Thus, if there is enough inflation to solve the horizon
problem the flatness problem is also be solved.

Now consider the opposite limit, large curvature,
$S_{\rm curv} \la \mpl^3 /{\cal M}^3$.  In this circumstance
the curvature term becomes comparable to the radiation-energy
density term
at a time when they are both larger than the vacuum-energy term.
If the Universe is positively curved this is very bad,
as the Universe will recollapse before can inflate,
and thus the flatness and horizon problems cannot be solved.
(There is a way out; in some models of inflation, e.g.,
chaotic inflation \cite{linde}, the Universe begins
vacuum dominated, i.e., $\rho \sim V(\phi )
\sim \mpl^4$, and inflating, and this problem never arises.)

For $k < 0$, the curvature term begins
to dominate the right-hand side of the Friedmann equation at a temperature
$T_{\rm curv} \sim S_{\rm curv}^{-1/3} \mpl$; the Universe
undergoes a period of ``free expansion,'' and $\Omega$
approaches zero.  When the temperature
reaches $T_{\rm inflate} \sim S_{\rm curv}^{1/3}({\cal M}/\mpl )^2
\mpl$, vacuum energy begins to dominate the right-hand
side of the Friedmann equation
and inflation begins.  The requirement on $N$ to solve the horizon
problem is precisely as before, cf. Eq.~(\ref{eq:nn}).

Now consider the flatness problem.  After inflation, when
the vacuum energy has been converted to radiation, the
entropy per comoving volume has
increased by a factor of $e^{3N}({\cal M}/T_{\rm inflate})^3
\sim e^{3N}(\mpl /{\cal M})^3 /S_{\rm curv}$.  Thus
the final entropy contained within a curvature radius is
\begin{equation}
S_{\rm curv} ({\rm final}) \sim e^{3N}(\mpl /{\cal M})^3.
\end{equation}
Note that the final entropy within a curvature radius does
not depend upon the initial value.
The condition that $S_{\rm curv}({\rm final})$ be
greater than about $10^{88}$ is the very same condition for solving
the horizon problem,
\begin{equation}
N \ga 68 + \ln ({\cal M}/\mpl ).
\end{equation}

(One might worry that the number of e-foldings of inflation
$N$, which is determined by the time it takes the
inflaton to roll to the minimum of its potential,
could depend upon the size of the curvature term
since it influences the expansion rate; it does not
in any significant way.  For a given
model of inflation, specified by the scalar potential, the
amount of inflation is, to within a few e-foldings,
independent of the size of the curvature term \cite{steigman}.)

To recapitulate, in the case of negative curvature sixty or so
e-foldings of inflation serves to solve the horizon and flatness
problems, regardless of the size of the initial curvature radius.
In the case of positive curvature, for many models of inflation
the Universe will only survive long enough to inflate if
$S_{\rm curv}$ is larger than $\mpl^3/{\cal M}^3$, which
is of the order of $10^{15}$ for the typical energy scale of inflation,
${\cal M} \sim 10^{15}\GeV$.

If one thought that the Universe began in an initial state
as simple and special as an FRW model this would be a little disturbing.
However, one expects that a generic initial state
is highly inhomogeneous, with regions of both negative and positive curvature.
Many, though not all,
of the positively curved regions would simple collapse to
form black holes, while {\it any} sufficiently large region of
negative curvature would undergo inflation and grow to a
size large enough to encompass all that we see today.
(Sufficiently large means much larger than the inverse
of the local expansion rate; see Refs.~\cite{inhomo}.)  In the
context of a generic beginning for the classical epoch it
is not very significant that only the negatively curved
regions are certain to undergo inflation.  In sharp contrast,
in the case of an adiabatic scenario, only those negatively curved
regions that contain an entropy greater than $10^{88}$
can ever be suitable to house our present Hubble volume.

\section{Variable Planck Mass Cannot Solve the Horizon Problem}
\subsection{The horizon problem restated}
\reset

We have seen in our previous discussion that an adiabatic scenario
cannot address the flatness problem (unless $S_{\rm curv}\ga
10^{88}$ the universe will not be as flat as ours is today).
However, one might well ask whether it is possible
to solve just the horizon problem in the post-Planckian era
without entropy production.  In this Section we show that in a wide
class of gravity theories this cannot be done, provided only that we
assume that that the various contributions to the energy density are
positive.  In particular, we rule out proposed adiabatic solutions based on a
time-varying Planck mass in the context of scalar-tensor
theories of gravity \cite{zee,soft,janna}.

For constant $\mpl$, it is easy to see that entropy production is
required to solve the horizon problem.  For adiabatic expansion, $S_1=S_0$
and Eq.~(\ref{eq:hsc}) requires $T_1 <10^2 T_0$ and thus $R_1 >10^{-2}
R_0$.  Since time $t_1$ was taken to be during the radiation-dominated
era, which began when $R\sim 10^{-4}R_0$, this inequality can only be
satisfied if the scale factor was decreasing at early times,
``bounced,'' and began increasing.  However, this cannot be, since in
the absence of entropy production
any FRW model that is now expanding must always have been
expanding, provided only that all energy densities are positive.

{}From Eq. (\ref{eq:hsc}) it might seem that a
decreasing Planck mass could drastically
alter this:  A larger Planck mass at early times would imply a weaker
effective gravitational constant.  This would lead to slower
expansion (at a given temperature), resulting in an
older Universe and a larger horizon
\cite{zee,soft,janna}. Specifically, Eq.~(\ref{eq:hsc})
would be satisfied without entropy production if
\begin{equation}  \label{eq:planck}
\mpl(t_1) \ga (10^{-2}\,T_1/T_0 )\,\mpp \sim 10^{30}T_1 ;
\end{equation}
where $\mpp = \mpl (t_0)= G_N^{-1/2}=1.22\times
10^{19}\GeV$ denotes the current value of the Planck
mass, and $\mpl (t) =G_N(t)^{-1/2}$ its value as
a function of time.
We now show that it is not possible to decrease the Planck mass
rapidly enough to satisfy Eq.~(\ref{eq:planck}).  Our strategy is to
focus on $T/\mpl$.  To reproduce the successful
predictions of primordial nucleosynthesis, the Planck mass must have
reached its present value by a temperature of $1\MeV$.  From
this and Eq.~(\ref{eq:planck}) it follows
that at time $t_1$ the value of $T/\mpl$ must
have been smaller than its value at nucleosynthesis
by at least a factor of $10^8$; in fact, as we shall see, adiabatic
expansion precludes $T/\mpl$ from increasing at all.

In describing a generic theory with a variable Planck mass
we shall represent the Planck mass squared by a Brans-Dicke
type field  $\Phi = \mpl^2$.   We
write the action in the form
\begin{equation}    \label{eq:action}
S = \int d^4x \sqrt{-g} \left[ -{\Phi\over 16\pi} {\cal R} +
   {\omega(\Phi) \over 16 \pi \Phi} \partial_\mu \Phi\partial^\mu \Phi
     - V(\Phi) + {\cal L}_{\rm matter}   \right] .
\end{equation}
The unusual form of the $\Phi$
kinetic energy term is not essential; it can be put in the standard form
by transforming to a field $\psi(\Phi )$ obeying $(d\psi/d\Phi)^2 =
\omega (\Phi )/(8\pi \Phi)$.  For constant $\omega$ and vanishing $V(\Phi)$
this reduces
to the Brans-Dicke theory.  We will assume
that both the matter energy density $\rho$ and $V(\Phi)$ are
non-negative (a negative potential would
lead to a negative cosmological constant).
It is also reasonable to require that $\omega(\Phi)$
be positive to avoid the instabilities and quantum mechanical
inconsistencies associated with negative kinetic and gradient
energy terms (actually, only the weaker condition $\omega \ge -3/2$ is
needed).  We do not consider the possibility of terms of second
or higher order in the curvature;
for the case of second-order terms, the theory
can be reformulated as Einstein gravity with
an additional field \cite{star} and an
analysis similar to ours can be applied.

This action leads to a Friedmann equation of the form
\begin{equation} \label{eq:f1}
  H^2 = {8\pi (\rho +V) \over 3 \Phi}  - H\left( {\dot\Phi \over \Phi}\right)
    + {\omega\over 6}\left({\dot\Phi \over \Phi}\right)^2 - {k \over R^2}.
\end{equation}
It is convenient to rewrite this as
\begin{equation}  \label{eq:f2}
 \left(H + {1\over 2} {\dot\Phi \over \Phi}\right)^2  =
  {8\pi (\rho +V )\over 3 \Phi}
    + {1\over 6}\left(\omega+ {3\over 2}\right)
        \left({\dot\Phi \over \Phi}\right)^2 - {k \over R^2}.
\end{equation}
The quantity appearing on the left-hand side of this equation is
\begin{equation}
H + {1\over 2} {\dot\Phi \over \Phi} = {d\over dt}\ln ( R \mpl) =
-{d\over dt}\ln ( T/\mpl );
\end{equation}
where the second equality follows from the assumption of adiabaticity.
Hence, during epochs where the right-hand side
of Eq.~(\ref{eq:f2}) is nonzero, the quantity $T/\mpl$ must evolve
monotonically.  For $k\le 0$, the right-hand side can
never be negative, and so the variation of $T/\mpl$ is always monotonic.
Since $T/\mpl$ has certainly been decreasing since the time of nucleosynthesis,
it must always have been doing so;
we thus have our result for an open or flat Universe.

The proof for a closed
Universe requires more work.  With $k>0$, the right-hand side of
Eq.~(\ref{eq:f2}) has no definite sign, and so
the Universe might have alternated
between eras of increasing and decreasing $T/\mpl$.
Suppose the current era of decreasing $T/\mpl$ began at some time $t_*$,
after the early era of rapid Planck mass variation and
before nucleosynthesis.  The vanishing of
the right-hand side of Eq.~(\ref{eq:f2})
at $t=t_*$ implies
\begin{equation}  \label{eq:closed}
  {k\over R^2(t_*)} \ge {8\pi \over 3} {\rho_{\rm rad}(t_*)\over \mpl^2(t_*)};
\end{equation}
These quantities can be related to the corresponding quantities at
the time of nucleosynthesis by using adiabaticity
and the fact that $T/\mpl$ has been decreasing since $t=t_*$ to obtain
\begin{equation}
{k\over R^2(t_{\rm BBN})} \ge
  { 8\pi \over 3} {\rho_{\rm rad}(t_{\rm BBN})\over \mpl^2 (t_{\rm BBN})}.
\end{equation}
This last inequality is false, since at the time of
nucleosynthesis the curvature term in the Friedmann equation was in fact
much smaller than the radiation energy density.  Hence,
the assumption of the existence of a time $t_*$ must be abandoned, and we
have proven our result.

\subsection{Conformal transformation to Einstein gravity}

It is instructive to use a conformal transformation to
rewrite our action with a constant Planck mass, but
time-varying particle masses.  In the conformal frame,
changing the sign of $d(T/\mpl )/dt$ is equivalent
to constructing a cosmological model that bounces.
The conformal transformation is accomplished
by defining a new metric
\begin{equation} \label{eq:conformal}
 \g_{\mu\nu} = {\Phi \over \mpp^2} g_{\mu\nu}.
\end{equation}
When expressed in terms of this metric and the corresponding Ricci
scalar $\tilde\R$, the action of Eq.~(\ref{eq:action})
becomes (after an integration by parts)
\begin{eqnarray}
{\tilde S} = \int d^4x \sqrt{-\g} \Biggl[
 -{\mpp^2\over 16\pi} {\tilde {\cal R}} & + &
   \left(\omega + {3\over 2}\right){ \mpp^2 \over 16 \pi \Phi^2}
\partial_\mu \Phi\partial^\mu \Phi  \nonumber \\
   & - & {\mpp^4 \over \Phi^2} V(\Phi)
       + {\mpp^4 \over \Phi^2}{\tilde {\cal L}}_{\rm matter}   \Biggr] .
\end{eqnarray}
(${{\tilde {\cal L}}}_{\rm matter}$ has
further $\Phi$-dependence because of the metric factors which it
contains.)  In this frame, $\Phi$ is no longer
the inverse of the effective gravitational constant,
but simply another matter field whose
contributions to the energy density and pressure
(assuming spatial homogeneity) are
\begin{equation} \label{eq:rhophi}
\rho_\Phi =  \left(\omega + {3\over 2}\right){ \mpp^2 \over 16 \pi \Phi^2}
\left({d\Phi\over d\tilde t}\right)^2  +{\mpp^4 \over \Phi^2} V(\Phi);
\end{equation}
\begin{equation}
p_\Phi =  \left(\omega + {3\over 2}\right){ \mpp^2 \over 16 \pi \Phi^2}
\left({d\Phi\over d\tilde t}\right)^2  - {\mpp^4 \over \Phi^2} V(\Phi).
\end{equation}
The time and scale factor for the
transformed metric are related to those for the original
metric by $d\tilde t/dt =
\sqrt{\Phi/\mpp^2}$ and $\tilde R =\sqrt{\Phi/\mpp^2} \,R$, so the
Hubble parameter for the transformed metric is
\begin{equation} \label{eq:ftilde}
\tilde H \equiv {1\over \tilde R}{d\tilde R \over d\tilde t}
   = {\mpp\over \sqrt{\Phi}}
 \left( H + {1\over 2} {\dot \Phi \over \Phi} \right) .
\end{equation}
We recognize Eq.~(\ref{eq:f2}) as the Friedmann equation,
in standard form, in the new frame.  Further,
$\tilde H$ is, up to to a positive numerical factor,
equal to $-d\ln (T/\mpl )/dt$.  Thus, to have the present era of
decreasing $T/\mpl$ preceded by a period in which $T/\mpl$ increased,
$\tilde R$ would have to first decrease and
then increase.   However,
an adiabatically expanding FRW Universe with positive energy density
cannot undergo such a bounce, as can be shown by a simple modification
of the arguments we have given above.

The conformal transformation allows us to close a small loophole.
In showing that an adiabatic solution to the horizon
problem requires $T/\mpl$ to increase, we used the bound $H_1
\ga T_1^2/\mpl$.  One might worry about the potentially
negative contribution of the $(\dot \Phi
/\Phi )H$ term in Eq.~(\ref{eq:f1}), though this term
is expected to be positive since $\mpl$ is decreasing.
In any case, it is simple to show that
$H_1/\sqrt{8\pi \rho_{\rm rad}/3\mpl^2} \ge [1 + (3/2\omega)]^{-1/2}$
for $k\le 0$.  Thus, a problem arises only if
$\omega$ is very small {\it and} $\mpl$ is increasing.
We could start with Eq.~(\ref{eq:horcondx}) and redo our analysis;
it is simpler to work in the conformal frame where we
have an ordinary FRW model, with an additional energy term,
cf.~Eq.~(\ref{eq:rhophi}).  Here it is easy to see that there cannot be
an adiabatic solution to the horizon problem.  Since the existence
of a horizon problem is independent of frame, we have our result.

\section{Concluding Remarks}
\reset

We have shown that a truly satisfactory dynamical resolution of the
horizon and flatness problems associated with
the standard cosmology requires both
superluminal (or very close to it)
expansion and massive entropy production.
In general, we have shown that an adiabatic scenario cannot solve the horizon
problem unless the Universe was very flat to begin with.
Further, for a large class of adiabatic scenarios, those based on scalar-tensor
theories, we have shown directly how adiabaticity by itself precludes
a solution to the horizon problem regardless of how flat the Universe is.

Because both superluminal expansion and entropy
production are necessary to solve in a satisfactory manner the horizon
and flatness problems
and because they are the two generic features of
all inflationary models, it is suggestive to conclude that inflation,
or something very similar, provides the only dynamical solution to these
vexing cosmological problems.

\vskip 1.5cm

\noindent We thank J.~Levin and K.~Freese for extensive discussions.
This work was supported in part by the DOE (at Chicago, Columbia,
and Fermilab) and by NASA through NAGW-2381 (at Fermilab).

\end{document}